\documentclass[12pt]{article}
\usepackage{amsmath}
\usepackage{graphicx,psfrag,epsf}
\usepackage{enumerate}
\usepackage{natbib}
\usepackage{url} % not crucial - just used below for the URL 
\usepackage{hyperref}
\hypersetup{
	colorlinks,
	citecolor=green,
	filecolor=black,
	linkcolor=black,
	urlcolor=blue
}

\newcommand{\blind}{1} 

\usepackage{color}
 
\usepackage{amssymb,bm} 
\usepackage{mathrsfs}

\newcommand{\x}{\mathbf{x}}

\renewcommand{\c}{\mathbf{c}}
\newcommand{\C}{\mathbf{C}}

\renewcommand{\t}{^\top}

\renewcommand{\i}{^{-1}}

\newcommand{\E}[1]{\mathbb{E}\left[#1\right]}

\newcommand{\V}[1]{\mathbb{V}\mathrm{ar}\left[#1\right]}
\newcommand{\I}{\mathbb{I}}

\newcommand{\argmax}{\operatornamewithlimits{arg\,max}} 

\newcommand{\cond}{\mspace{3mu}{|}\mspace{3mu}}

\begin{document}

\bibliographystyle{apalike}

\def\spacingset#1{\renewcommand{\baselinestretch}%
{#1}\small\normalsize} \spacingset{1}

%%%%%%%%%%%%%%%%%%%%%%%%%%%%%%%%%%%%%%%%%%%%%%%%%%%%%%%%%%%%%%%%%%%%%%%%%%%%%%

\if1\blind
{
 \title{\bf Gaussian Processes for Individualized Continuous Treatment Rule Estimation}
 \author{Pavel Shvechikov 
% 	\\
 and Evgeniy Riabenko\thanks{ The reported study was funded by RFBR, according to the research project No. 16-07-01192 А.}\hspace{.2cm}\\
 Department of Computer Science, Higher School of Economics}
 \maketitle
} \fi

\if0\blind
{
 \bigskip
 \bigskip
 \bigskip
 \begin{center}
 {\LARGE\bf Gaussian Processes for Individualized Continuous Treatment Rule Estimation}
\end{center}
 \medskip
} \fi

\bigskip
\begin{abstract}
Individualized treatment rule (ITR) recommends treatment on the basis of individual patient characteristics and the previous history of applied treatments and their outcomes. Despite the fact there are many ways to estimate ITR with binary treatment, algorithms for continuous treatment have only just started to emerge. We propose a novel approach to continuous ITR estimation based on explicit modelling of uncertainty in the subject’s outcome as well as direct estimation of the mean outcome using gaussian process regression. Our method incorporates two intuitively appealing properties -- it is more inclined to give a treatment with the outcome of higher expected value and lower variance. Experiments show that this direct incorporation of the uncertainty into ITR estimation process allows to select better treatment than standard indirect approach that just models the average. Compared to the competitors (including OWL), the proposed method shows improved performance in terms of value function maximization, has better interpretability, and could be easier generalized to multiple interdependent continuous treatments setting.
\end{abstract}

\noindent%

\vfill

\newpage
\spacingset{1.45} % DON'T change the spacing!

\section{Introduction}
\label{sec:intro}

Many clinician studies conclude that reactions to a fixed treatment could vary significantly among individuals in a population \citep{ishigooka2000olanzapine, havlir2011timing, bolondi2012heterogeneity}. Such heterogeneity of treatment effects \citep{kravitz2004evidence} raises a question about the applicability of this "one size fits all" approach to clinical practice and leads naturally to personalized medicine paradigm -- the idea of healthcare improvement as a result of tailoring treatments on the basis of patient individual characteristics. 

\subsection{Treatment rules}
One of the crucial concepts in personalized medicine research is \emph{individualized treatment rule} (ITR) (aka individualized treatment \emph{regime}) which is usually defined as a mapping from measured patient characteristics to recommended treatment \citep{laber2017evaluating}. An obvious goal of ITR construction is a maximization of patients' response to treatment averaged over all individuals in a population. This goal is sometimes referred to as expected reward in a population or \emph{value function}. ITR estimation is a procedure which uses data from either observational or randomized clinician studies to fulfill this goal.

Personalized medicine often comes in one of the two forms \citep{lipkovich2017tutorial}: either identification of the subgroups of patients who benefit from a given treatment, or identification of the optimal treatment for a specific patient. The latter form (which we also adhere to in this work) could be further subdivided by the approach to ITR estimation:
\begin{enumerate}\itemsep0em 
	\item \emph{Indirect approach} usually consists of a two-stage procedure. First, modeling of either conditional mean outcomes or contrasts between mean outcomes takes place. Second, the ITR is derived as the rule assigning the treatments that maximize the model of the conditional mean outcome.
	
	\item \textit{Direct approach} approximates expected reward in a population with its sample estimate as a function of ITR. Maximization of this expected reward estimate with regard to parameters of ITR yields the desired treatment rule. 
\end{enumerate}

Specific ITR estimation methods differ a lot depending on whether the treatment is discrete or continuous. In the latter case ITR is also called \emph{individualized dose rule} (IDR) and is of a particular practical importance because of its usage in therapies balancing drug dose between insufficient (not causing any effect) and excessive (causing severe adverse effects). One example of such dose administering therapies is thrombosis prevention using a drug called warfarin, which is shown to be more effective when prescribed on the basis of environmental and genetic patient factors \citep{international2009estimation}. Another illustrative example is a treatment of inflammatory bowel disease with thiopurines \citep{teml2007thiopurine}. 
These examples motivate a strong need in developing reliable and efficient methods for ITR estimation. 

There is a substantial amount of literature devoted to ITR estimation dealing with discrete (mostly binary) treatment using an indirect approach \citep{qian2011performance, moodie2014q, zhang2015c, tao2017adaptive} as well as a direct one \citep{zhao2012estimating, zhou2015residual}. However, despite its importance, as far as we know, there are very few methods dealing with IDR estimation. 

IDR estimation poses additional challenges because the set of admissible treatments has a cardinality of the continuum. There are infinitely many treatments to choose from, only a few of which might be actually present in the data. These difficulties yield trivial generalization of discrete methods to continuous treatment cases either impossible or intractable. 

One of the recent works devoted to IDR estimation \cite{Laber2015} follows a direct approach with the focus on policy interpretability, which is achieved through the use of decision trees with splits guided by complicated purity measure. This split criterion is based on the usual value function from discrete ITR with additional smoothing. The proposed smoothing is analogous to introducing a stochastic policy with probability mass concentrated around the deterministic policy assignments observed in the training data. An interesting peculiarity of the work is that only treatments observed in the training data are considered as a split candidates. 

Another recent approach to IDR estimation \citep{chen2016personalized} is devoted to developing an extension of O-learning algorithm \citep{zhao2012estimating}. It resembles in spirit the work of \cite{Laber2015} because of chosen \emph{direct} approach to IDR estimation and the presence of smoothing in value function estimate. However, unlike in \cite{Laber2015}, the smoothing is used only as an intermediate step required for optimization purposes, so that the annealing of the kernel bandwidth to zero gives precisely the same value function as is used for discrete ITR estimation. 

Specifically, \cite{chen2016personalized} introduced a regression based approach inspired by the \cite{zhao2012estimating} and SVM; the proposed algorithm comprises global optimization task tackled by DC procedure which is known to depend on initialization and could guarantee only local convergence.

All of the aforementioned algorithms follow the direct approach to IDR estimation consisting of maximizing mean conditional reward over individuals in a population (aka value function), but do not deal with variance of the patients' potential outcome, which might be crucial for a real world application. 

Consider a simple example when patient is told to choose between two mutually exclusive options: 
\begin{enumerate}\itemsep0em 
	\item Take a drug A with 0.99 chance of recovery (reward = 100) and 0.01 probability to acquire physical disability (reward = -1000). 
	\item Take a drug B with 0.85 chance of recovery (reward = 100) and 0.15 probability not to experience any effect (reward = 0).
\end{enumerate}

Mean reward in the first case is 89 while in the second one it is 85, suggesting that rational patient should always choose the first option. However, it is known that people are not rational and tend to averse losses \citep{kahneman1979prospect}. This common behavior was also shown to be true in a health domain \citep{attema2013prospect}. Therefore, it seems very natural to explicitly incorporate reasoning about mean and variance of a possible outcome into an ITR. Nevertheless, such reasoning is impossible when using \emph{direct} approach. 

\subsection{Indirect methods}

Our work aims to develop an \emph{indirect} approach to ITR estimation. \emph{Indirect} methods are usually criticized for potential overfitting during the first stage (fitting regression model to rewards) and for possible model misspecification.

In defense of \emph{indirect} methods, we argue that the concerns about misspecification could be reduced to a great extent through the use of a flexible model that could fit the data well \citep{taylor2015reader}. 

However, the more powerful model is, the more it is susceptible to overfitting. To cope with the overfitting problem on the level of model design, we have chosen to follow Bayesian paradigm, which allow to do natural regularization through the averaging over model parameters and is known to be free of overfitting problem \citep{bishop2006pattern}. Nevertheless, any Bayesian model becomes prone to overfitting as soon as optimization with regards to hyperparameters is involved (so-called type II maximum likelihood procedure). Careful treatment of overfitting requires either very small number of hyperparameters or fully Bayesian inference with Markov chain Monte Carlo methods \citep{neal1997monte, Gramacy2007}.

Another argument in favor of \emph{indirect} methods is their capacity to estimate the uncertainty of the expected outcome. Such estimation is impossible when following \emph{direct} approach but could be invaluable in practice, providing the means for balancing exploration and exploitation.

%Another argument in favor of indirect methods comes from an observation that 
General task of ITR estimation is similar to the classic reinforcement learning (RL) setting \citep{sutton1998reinforcement}: patient characteristics, their treatments and responses are direct analogs of state, action and reward concepts from RL literature. Moreover, \textit{direct} and \textit{indirect} approaches to ITR estimation in RL community are called value-based and policy-based approaches. Recently, it was shown \citep{o2016pgq, schulman2017equivalence} that these two approaches share much more in common than it was believed before. To some extent this finding reconciles two different ITR estimation approaches and gives rise to an unexplored idea of their combination.

The contributions of this work are summarized as follows. 
\begin{enumerate}\itemsep0em 
	\item We propose a new \emph{indirect} approach to IDR estimation that benefits from separation of uncertainty due to finite sample estimation and irreducible variance of patient outcomes. 
	\item We show how to reduce the hard task of continuous inversion to simple univariate global optimization problem which is well-known in active learning and Bayesian optimization literature. 
	\item We evaluate the proposed method on various simulation studies, confirming its superior performance for randomized dose-finding clinical trials. 
\end{enumerate}

\section{Proposed method}
We assume that train data comes from randomized dose trial and every training record consists of three major components: 
\begin{enumerate}\itemsep0em 
	\item patient covariates $ \c \in \mathbb{R}^p$;
	\item treatment doze $ a \in [0,1] $, selected at random within safe bounded dose range, which w.l.o.g. is assumed to be equal to $ [0, 1] $ interval;
	\item patient outcome $ r \in [r_{lo}, r_{hi}], ~ r_{lo}, r_{hi} \in \mathbb{R}$.
\end{enumerate}	
Note that we do not restrict the sign of outcomes but do require them to be bounded. We also introduce a~shorthand notation $ \x_i = [\c_i\t, ~ a_i]\t $ for a~concatenation of covariate vector and a treatment value.

As was mentioned previously, \emph{indirect} approach to IDR estimation requires building a~model of the outcome value given patient covariates and treatment doze. 
One of the models meeting all criteria of being sufficiently flexible, robust to overfitting due to Bayesian nature, and able to make probabilistic inference is Gaussian process (GP) regression model \citep{rasmussen2005GPM}, which is the basis of this work.

\subsection{Gaussian process framework} 
We proceed with minimal introduction to GP regression. For more thorough discussion one could refer to \cite{rasmussen2005GPM}. 

Gaussian process is a collection of random variables any finite number of which have joint Gaussian distribution. A GP is completely specified with its mean and covariance functions which we will denote as $ m(\x) $ and $ k(\x, \x') $ respectively. For clarity and without loss of generality \citep{rasmussen2005GPM} we will consider Gaussian process $ f $ with zero mean $ m(\x) = 0 $ and denote $ f(\x) \sim \mathcal{GP}(0,~ k(\x, \x'))$. 

We assume that we were given a training dataset $ \{(\x_i , y_i ) \mid i = 1,..., n\} $ that consists of pairs of object description $ \x_i $ and noisy observation of function realization, i.e. $y_i = f(\x_i) + \varepsilon$, where the noise $ \varepsilon \sim \mathcal{N}(0,\sigma_n^2 ) $.

Having specified a joint Gaussian prior over the test output $ f_* $ and noisy train outputs~$ \bf{y} $, the inference is carried out with conditioning on observed train outputs. This procedure allows to compute the mean and the variance of $ f_* $ posterior distribution given the test point $ \x_* $ using the following equations:
\begin{equation}
\begin{aligned} \label{eq:gp_predictions}
f_* &\sim \mathcal{N}(\bar f_*, ~ \V{f_*}) \\
\bar f_* &= \bm{k}_*\t \left(K + \sigma^2_n I\right)\i \bm{y} \\
\V{f_*} &= k(\x_*, \x_*) - \bm{k}_*\t \left(K + \sigma^2_n I\right)\i \bm{k}_*,
\end{aligned}
\end{equation}
where $ \bm{k}_*\t = (k(\x_*, \x_1), ..., k(\x_*, \x_n)) $ and covariance matrix $ K $ has entries $ K_{ij} = k(\x_i, \x_j) $. To reduce clutter we will subsequently denote $ \left(K + \sigma^2_n I\right)\i$ as $ \Lambda $. 

One of the main constitutes of GP flexibility is covariance function.
The choice of covariance function for particular application can be guided by Bayesian model selection \citep{rasmussen2005GPM}, however, for illustration purposes we have chosen probably the most commonly used squared exponential function with separate length-scales. We also extend this covariance function with the noise (aka nugget) term:
\begin{align*}
k(\x_p, \x_q) 
= \sigma_f^2 \exp \left\{ - \sum_{i=1}^p \dfrac{(x_{pi} - x_{qi})^2}{2\theta_i} \right\} + \sigma_n^2	\delta_{pq}, 
\end{align*}
where $ \theta_i $ is called length-scale, $ x_{pi} $ is a $ p $-th feature of $ i $-th object and $ \delta_{pq} $ is the Kronecker delta function. 

Note that all of the subsequent derivations could be replicated with other popular choices of covariance function (i.e. the Mat\'{e}rn class) and are not unique to squared exponential covariance function.

\subsection{Method formulation}

Motivation of the proposed method is as follows. 
To solve the prescription problem for a~new patient with characteristics $ \c_* $, doctor, given the history of previously seen patients $ \{(\c_i, a_i, r_i)\}_{i=1}^n $, have to simultaneously solve two conflicting tasks:
\begin{enumerate} \itemsep0em 
	\item Doctor has to be \textbf{confident} about potential dose effect and this confidence should be based on previous cases. 
	The more similar are patient characteristics $ \c_* $ to the ones from previous practice $ \c_i $, the more doctor is sure that giving the same $a_i $ to patient with $ \c_* $ will result in $r_i$. 
	
	\item Doctor has to prescirbe \textbf{the best} treatment possible. The bigger is $ r_i $, the more doctor wants $ a_* $ to be similar to $ a_i $.
\end{enumerate}

These two intuitive observations coupled with the notion of loss aversion in health domain \citep{attema2013prospect} motivate prescription of the dose which maximizes probabilistic lower surface for the potential outcome. That is, in terms of mean and variance of posterior distribution over $ f_* $
\begin{equation} \label{eq:lbcs}
a_* = \argmax_a ~~ \bar f_* - s \sqrt{\V{f_* }},
\end{equation} 
where $ s $ is a scaling coefficient controlling the penalty for the uncertainty in the outcome. 

We argue that this simple heuristic, which hereinafter will be referred to as LCSL (lower confidence surface learning):
\begin{enumerate}\itemsep0em 
	\item is a strong IDR competitive with the previously proposed approaches;
	\item provides the means to account for loss aversion and uncertainty;
	\item provides the means to perform adaptive trial design;
	\item allows to explicitly address an exploration-exploitation tradeoff.
\end{enumerate}

\subsection{Modelling correlation with exponential power family}

Formula (\ref{eq:lbcs}) does not gives us a method to perform maximization over treatment domain. 
So we need to design a constructive way to carry out such an inversion.

Lets rewrite each of the terms in (\ref{eq:lbcs}) separately assuming that covariance function is squared exponential with separate length-scales. 
For the sake of brevity we omit length-scales in following derivations
$$
k(\x_p, \x_q) 
= \sigma_f^2 \exp \left\{ -\dfrac{||\x_p - \x_q||^2_2}{2} \right\} 
= \sigma_f^2 \exp \left\{ -\dfrac{||\c_p - \c_q||^2_2}{2} \right\} \exp \left\{ -\dfrac{||a_p - a_q||^2_2}{2} \right\}
$$
Lets denote
$$
 K(\c_*) = \sigma_f^2 \cdot \mathrm{diag} \left( \exp \left\{ -\frac{||\c_* - \c_1||^2_2}{2} \right\}, ~\dots~, \exp \left\{ -\frac{||\c_* - \c_n||^2_2}{2} \right\} \right)
%	\begin{pmatrix}
%	\ & \cdots & 	\mbox{\large 0} \\
%	\vdots & \ddots & \vdots \\
%	\mbox{\large 0} & \cdots & 
%	\end{pmatrix}
$$
Then one could write $ \bar f_* $ in a form of a linear combination of exponential terms where $ \alpha_i $ depends only on the train data and known covariates $\c_* $ of test object:
\begin{align*}
\bar f_* &= \begin{pmatrix}
\exp \left\{ -\dfrac{||a - a_1||^2}{2} \right\} & , ~\dots~, & \exp \left\{ -\dfrac{||a - a_n||^2}{2} \right\} 
\end{pmatrix}K(\c_*)\t \Lambda \bm{y} \\
&= \sum_i \alpha_i	\exp \left\{ -\dfrac{||a - a_i||^2}{2} \right\}, 
\end{align*}
where $ \alpha_i = \left[K(\c_*)\t \Lambda \bm{y} \right]_i $.

The same transformation could be done for $ \V{f_*} $:
\begin{align*}
\V{f_*} 
&= k(\x_*, \x_*) - \bm{k}_*\t \Lambda \bm{k}_* \\
&= k (\x_*, \x_*) - \sum_{i,j} \gamma_{ij} \exp \left\{ -\frac{||a - a_i||^2 + ||a - a_j||^2}{2} \right\} 
\end{align*}
Here $ \gamma_{ij} = \sigma_f^4 \Lambda_{ij} \exp \left\{ -\frac{||\c_* - \c_i||^2 + ||\c_* - \c_j||^2}{2} \right\} $
and $\forall \, x_* ~ k( \x_*, \x_*) = \sigma_f^2 $, so neither $ \beta_{ij} $ nor $ k( x_*, x_*) $ depends on unknown $ a $.

To sum up, our optimization task could be simplistically formulated as follows
\begin{equation} \label{eq:objective}
\max_a ~	\sum_i \alpha_i	\exp \left\{ -\frac{(a - a_i)^2}{2} \right\} - s \sqrt{ k( \x_*, \x_*) - \sum_{i,j} \gamma_{ij} \exp \left\{ -\frac{(a - a_i)^2 + (a - a_j)^2}{2} \right\} } 
\end{equation}

\subsection{Global optimization}
\label{sec:glob_opt}
As is easily seen, the objective (\ref{eq:objective}) is a sum of unnormalized exponential functions with coefficients $ \alpha_i, \gamma_{ij} $ of arbitrary sign. Thus, to allow for explicit maximization of this nonconvex objective, one need to employ global optimization methods. 

Even though the task (\ref{eq:objective}) is a very hard one, it is surely not a new one -- problems of this type are inevitable in Bayesian optimization \citep{brochu2010tutorial}, arising during maximization of the so-called acquisition function. 

There are many methods to solve global optimization problem of the form (\ref{eq:objective}), i.e. DIRECT \citep{jones1993lipschitzian}, which is derivative free and deterministic. In this work we adhere to a simpler yet efficient two-step approach to global optimization that is often implemented in Bayesian optimization frameworks \citep{bayesopt2014, spearmint2014}. First, seeds are generated -- either uniformly at random, in a form of fixed step grid or as a Sobol' sequence \citep{sobol1967distribution}. Second, function is evaluated at each seed and (optionally) local optimization method is started with each seed as an initial point. Frequent choice of such local optimization algorithm is L-BFGS-B \citep{Zhu1997}.

\subsection{Interpretability of the model}

Flexibility of any model usually comes with the price of reduced interpretability which is often of particular interest for practical applications. 
However, as is evident from formula (\ref{eq:gp_predictions}), mean predictions of GP regression are additive with respect to covariance terms $ \bm k_* $. 
This means that, given test object $ \x_* $, one could argue which objects from the train dataset influence mean prediction the most based on the components of covariance vector $ \bm k_* $ and coefficients $ [\Lambda \bm y]_i $. The same reasoning holds for variance interpretation with the only difference of quadratic dependence replacing the linear one.

Additionally, the learned values of length-scales could serve as indicators of feature importances, giving rise to the so-called automatic relevance determination (ARD) effect.

These simple model interpretations coupled with the recent work devoted to the tradeoff between accuracy and interpretability \citep{plate1999accuracy} along with possible interpretability increase due to additive GP modelling \citep{duvenaud2011additive} allow to get even closer to interpretability level of linear and tree-based models.

\subsection{Applicability to observational dose trials}

The requirement of training data being obtained from randomized study is by no means necessary. Under usual assumptions in casual inference literature \citep{hernan2017causal} the proposed method could be seen as a model of counterfactual mean outcome. These assumptions are as follows:
\begin{enumerate}\itemsep0em 
	\item Conditional exchangeability (aka strong ignorability) -- potential outcomes $ R^a $ are conditionally independent of $ A $ given covariates $ C $, i.e. $ \forall a, ~ R^a \perp A \cond C $.
	\item Positivity -- there is nonzero probability of assigning any of the treatment levels to any of possible objects, i.e. $ \forall C \colon p(C) > 0, ~ \forall a ~~ p(a \cond C) > 0 ~ $.
	\item Consistency -- each observed outcome is precisely the one that would have been observed under given treatment level, i.e. $ R = R^A $.
\end{enumerate}

In case these assumptions hold true, structural model $ \E{R^a \cond C = c} $ of counterfactual outcome $ R^a $ is equal to conditional mean model $ \E{R \cond C = c, A = a} $. This imply that the proposed method falls under umbrella of \textit{outcome regression} causal inference models.

\section{Numerical results}

All experiments, unless stated otherwise, were conducted as follows. Train and test sample generation was repeated 50 times for each combination of scenario, train sample size, algorithm and variance penalty factor. Test sample size was fixed at the value of 1000. Rewards in train dataset were scaled to fit into $ [0,1] $. For each train dataset we fit a Gaussian process regression that was allowed to internally restart optimization of marginal likelihood $ 10 $ times. The model maximizing marginal likelihood across these $ 10 $ restarts is chosen for subsequent evaluation on test data. Approximate model performance is obtained by plugging predicted optimal dose for test objects into the true underlying state-action value function $ Q(\C, A) $. Method quality is reported in terms of $ \hat{ \mathcal{V}}(f) $, which is defined as approximate model performance averaged over all data generations. 

We have empirically observed that the choice of a particular global optimization method out of those described in section \ref{sec:glob_opt} does not change performance much. Thus, we report all of the results using the simplest one -- fixed step grid evaluation with number of grid point fixed at 50 for all of the simulation scenarios and settings. 

In this section we refer to our method as LCSL.$ X $, where $ X $ specifies the value of variance penalty (\ref{eq:objective}) by the equation $ s = \Phi^{-1}(X / 100) $, where $ \Phi $ is normal CDF. 

All our experiments are built on top of Gaussian process framework \cite{gpy2014}.

\subsection{Simulation study}

In this section we first present results of two simulation experiments which show that for simple scenarios our method works remarkably well.

We denote $ \C $ for patient covariates matrix, i.e. $ \C = [\c_1, ..., \c_n]\t $, where $ \c_i $ is a column vector of $ i $-th patient characteristics; $ A $ for a column vector of treatments, i.e. $ A = [a_1, ..., a_n]\t $; $ C_i $ for an $ i $-th column of $ \C $ matrix.

\begin{figure}[h!]
	\centering
	\includegraphics[width=\linewidth]{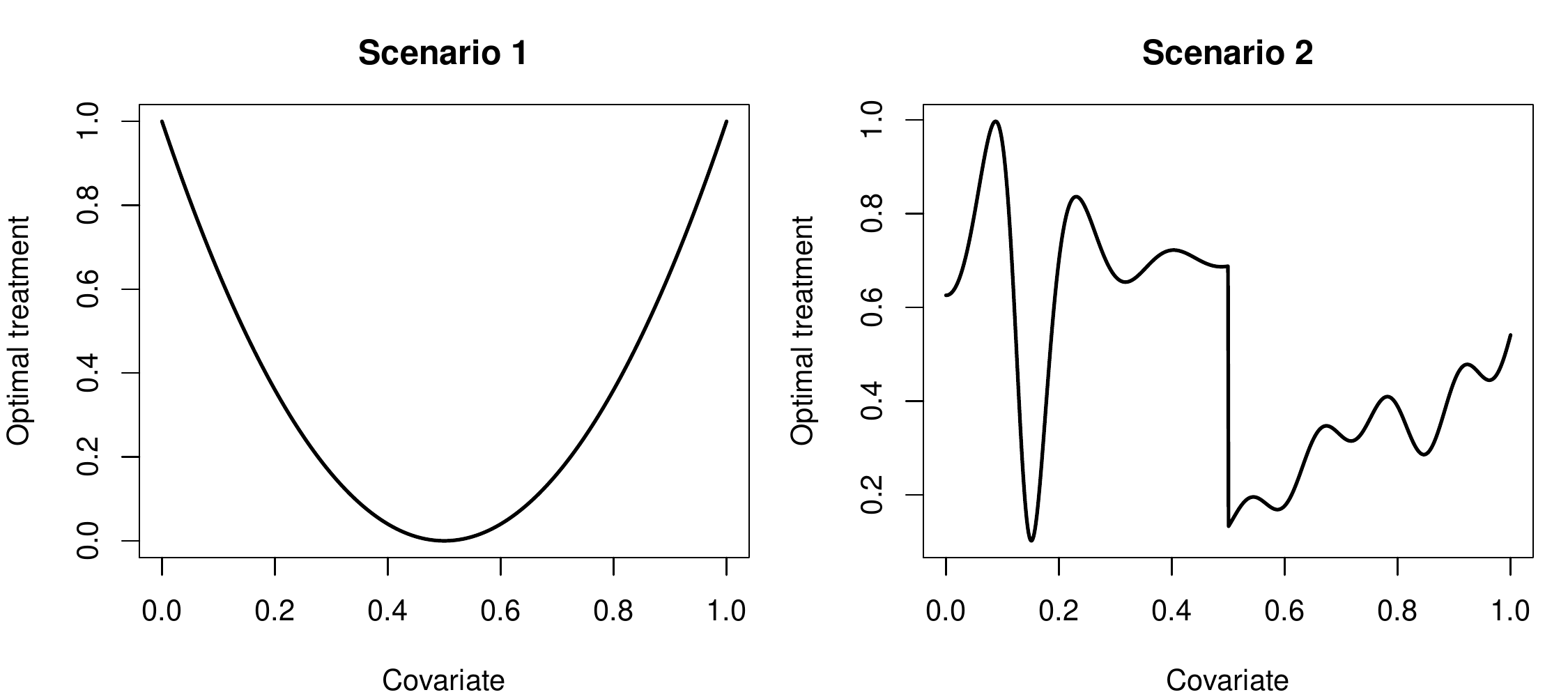}
	\caption{Illustration of $ f^{opt}(\C) $ from each of the proposed scenarios 1 and 2. }
	\label{fig:scenarios}
\end{figure}

Our first scenario simulates optimal treatment dependence on single covariate in a form of parabola (figure \ref{fig:scenarios}, left):
\begin{gather*}
	R ~ \sim ~ \mathcal{N}( Q_{1}(\C, A), 0.01), \qquad A ~ \sim ~ \mathcal{U}(0,1), \qquad \c_i ~ \sim ~ \mathcal{U}(0,1) \hfill \\
	Q_{1}(\C, A) = - 100 \cdot (f^{opt}_{1}(\C) - A)^2 \\
	f^{opt}_{1}(\C) = 4 \cdot (C_1 - 0.5)^2
\end{gather*}

Our second scenario is a more complex one -- it simulates optimal treatment assignment as a highly nonlinear function with abrupt change in the middle of covariate space and local periodicity (figure \ref{fig:scenarios}, right) :
\begin{gather*}
R ~ \sim ~ \mathcal{N}( Q_{2s}(\C, A), 0.01), \qquad A ~ \sim ~ \mathcal{U}(0,1), \qquad \c_i ~ \sim ~ \mathcal{U}(0,1) \hfill \\
Q_{2}(\C, A) = - 100 \cdot (f^{opt}_{2}(\C) - A)^2 \\
f^{runge}(x) = 	\cos(3 \pi (4x - 0.3)) / (1 + 25 (4x - 0.55) ^2) \\
f^{step}(x) = 0.1 x \cdot (10 + \sin(20x) + \sin(50x) ) - 1.3 \\
f^{opt}_{2}(\C) = (f^{runge}(C_1) + f^{step}(C_1) \cdot \I(C_1 > 0.5)) / 1.5 - 0.7
\end{gather*}

Because true $ f^{opt}(\C) $ is nonlinear in both scenarios, we don't include linear OWL (L-OWL) from \cite{chen2016personalized} into comparison. Kernel OWL (K-OWL) results are obtained by reusing the code from the supplementary material of \cite{chen2016personalized}. 

Simulation results (table \ref{tbl:shvechikov1}, \ref{tbl:shvechikov2}) show that the proposed method works extremely well in both scenarios, showing unimprovable results for the first scenario.

\begin{table}[h!]
	\centering
	\caption{Mean and std of $ \hat{\mathcal{V}}(f) $, Scenario 1}
	\begin{tabular}{|c|c|c|}
		\hline
		Sample size & LCSL.95 & K-OWL \\
		\hline
		50 & \textbf{-0.01 (0.00)} & -2.30 (1.37) \\
		100 & \textbf{0.00 (0.00)} & -0.93 (0.50) \\
		200 & \textbf{0.00 (0.00)} & -0.42 (0.18) \\
		400 & \textbf{0.00 (0.00)} & -0.23 (0.08) \\
		800 & \textbf{0.00 (0.00)} & -0.17 (0.06) \\
		\hline
	\end{tabular}
	\label{tbl:shvechikov1}
\end{table}

\begin{table}[h!]
	\centering
	\caption{Mean $ \hat{\mathcal{V}}(f) $, Scenario 2}
	\begin{tabular}{|c|c|c|}
		\hline
		Sample size & LCSL.95 & K-OWL \\
		\hline
		50 & \textbf{-3.78 (1.39)} & -3.92 (0.63) \\
		100 & \textbf{-2.19 (1.31)} & -3.63 (0.37) \\
		200 & \textbf{-0.40 (0.26)} & -3.37 (0.23) \\
		400 & \textbf{-0.22 (0.21)} & -3.21 (0.20) \\
		800 & \textbf{-0.10 (0.06)} & -3.06 (0.19) \\
		\hline
	\end{tabular}
	\label{tbl:shvechikov2}
\end{table}

We proceed with results of LCSL on several more complex and higher dimensional simulations first introduced in \cite{chen2016personalized}. In particular, we evaluate and compare our method on three scenarios, first two of which simulate randomized trial and the last one simulates observational trial.

Scenario 3, randomized trial:
%\begin{equation*}
\begin{gather*} 
R ~ \sim ~ \mathcal{N}( Q_3(\C, A), 1), \qquad A ~ \sim ~ \mathcal{U}(0,2), \qquad \c_i ~ \sim ~ \mathcal{U}(-1,1)^{30} \hfill \\
Q_3(\C, A) = 8 + 4 C_1 - 2C_2 - 2C_3 - 25 \cdot (f^{opt}_3(\C) - A)^2 \\
%	8 + 4 \cos(2\pi C_2) - 2 C_4 - 8 C^3_5 - 15 \cdot |f_{opt}(C)-A|\\ 
f^{opt}_3(\C) = 1 + 0.5 C_1 + 0.5 C_2
\end{gather*}
%\end{equation*} 

\bigskip
Scenario 4, randomized trial:
%\begin{equation*}
\begin{gather*}
R ~ \sim ~ \mathcal{N}(Q_4(\C, A), 1), \qquad A ~ \sim ~ \mathcal{U}(0,2), \qquad \c_i ~ \sim ~ \mathcal{U}(-1, 1)^{10} \\
Q_4(\C, A) = 8 + 4 \cos(2\pi C_2) - 2 C_4 - 8 C^3_5 - 15 \cdot |f^{opt}_4(\C)-A|\\ 
f^{opt}_4(\C) = 0.6 \cdot \I(|C_1| \geq 0.5) + C^2_4 + 0.5 \log(|C_7|+1) 
\end{gather*}
%\end{equation*} 

\bigskip

Scenario 5, observational trial:
%\begin{equation*}
\begin{gather*}
R ~\sim \mathcal{N}(Q_5(\C, A), 1), \qquad A \sim \text{Trunc}\mathcal{N}( f^{opt}_5(\C), 0, 2, 0.5), \qquad \c_i ~ \sim ~ \mathcal{U}(-1,1)^{10} \\
Q_5(\C, A) \equiv Q_4(\C, A)
% 8 + 4 \cos(2\pi C_2) - 2 C_4 - 8 C^3_5 - 15 \cdot |f^{opt}_4(\C)-A|\\ 
\\
f^{opt}_5(\C) \equiv f^{opt}_4(\C)
% 0.6 \cdot \I(|C_1| \geq 0.5) + C^2_4 + 0.5 \log(|C_7|+1) 
\end{gather*}
%\end{equation*} 

Results of randomized study simulation (tables \ref{tbl:chen1}, \ref{tbl:chen2}) allow to conclude that our method outperforms every other competitor (often by a large margin) in a randomized trial setting. Such superiority originates from the randomization of treatment assignment, which increases the accuracy of global response surface model and, as a result, improves quality of treatment recommendations.

\begin{table}[h!]
	\centering
	\caption{Scenario 3: Mean and (std) of $ \hat{\mathcal{V}}(f) $. Results with $ ^* $ are from \protect\cite{chen2016personalized}}
	\begin{tabular}{|c|c|c|c|c|c|}
		\hline
		Sample size & LCSL.95 & K-OWL* & L-OWL*& SVR* & Lasso* \\
		\hline
		50 & \textbf{7.65 (0.92)} & 4.78 (0.48) & 4.83 (1.40) & -12.21 (7.53) & -15.62 (6.61) \\
		100 & \textbf{7.87 (0.11)} & 5.69 (0.40) & 5.39 (0.93) & -2.57 (6.34) & -9.76 (5.21) \\
		200 & \textbf{7.93 (0.11)} & 6.68 (0.26) & 6.85 (0.34) & 3.46 (1.97) & -1.91 (4.65) \\
		400 & \textbf{7.97 (0.09)} & 7.28 (0.15) & 7.41 (0.14) & 6.13 (0.47) & 7.95 (0.01) \\
		800 & \textbf{7.99 (0.09)} & 7.54 (0.08) & 7.67 (0.08) & 7.36 (0.12) & 7.97 (0.01) \\
		\hline
	\end{tabular}
	\label{tbl:chen1}
\end{table}

\begin{table}[h!]
	\centering
	\caption{Scenario 4: Mean and (std) of $ \hat{\mathcal{V}}(f) $. Results with $ ^* $ are from \protect\cite{chen2016personalized}}
	\begin{tabular}{|c|c|c|c|c|c|}
		\hline
		Sample size & LCSL.95 & K-OWL* & L-OWL* & SVR* & Lasso* \\
		\hline
		50 & \textbf{2.19 (1.97)} & 2.00 (0.29) & 1.16 (0.71) & -1.96 (1.70) & -5.58 (2.27) \\
		100 & \textbf{4.46 (0.76)} & 2.19 (0.43) & 1.57 (0.52) & 0.24 (1.42) & -4.12 (2.17) \\
		200 & \textbf{5.30 (0.24)} & 2.84 (0.37) & 2.02 (0.30) & 2.01 (0.84) & -3.37 (1.38) \\
		400 & \textbf{5.58 (0.20)} & 3.69 (0.27) & 2.30 (0.18) & 3.47 (0.37) & -0.92 (3.12) \\
		800 & \textbf{5.75 (0.21)} & 4.41 (0.19) & 2.49 (0.10) & 4.35 (0.19) & 1.92 (0.69) \\
		\hline
	\end{tabular}
	\label{tbl:chen2}
\end{table}

\begin{table}[h!]
	\centering
	\caption{Scenario 5: Mean and (std) of $ \hat{\mathcal{V}}(f) $. Results with $ ^* $ are from \protect\cite{chen2016personalized}}
	\begin{tabular}{|c|c|c|c|c|c|}
		\hline
		Sample size & LCSL.95 & K-OWL & K-OWL*  & SVR* \\
		\hline      
		50 & -1.15 (2.55) & \textbf{1.88 (0.60)} & NA  & NA \\
		100 & -0.45 (3.91) & \textbf{2.43 (0.42)} & NA  & NA \\
		200 & 2.23 (3.94) & 3.27 (0.32) & \textbf{3.29 (0.28)} & -0.95 (1.57) \\
		400 & \textbf{4.26 (3.17)} & 4.22 (0.28) & NA  & NA \\
		800 & \textbf{5.56 (0.65)} & 4.95 (0.21) & 4.91 (0.14) & 3.04 (0.52) \\
		\hline
	\end{tabular}
	\label{tbl:chen4}
\end{table}

However, for simulation of observational study (table \ref{tbl:chen4}) we observe that out method yields superior results only when given enough data. 
Small samples do not allow to properly model response surface because all of the points are already concentrated in near optimal regions, giving insufficient information about treatment dependence on covariates.
This effect could be mitigated to some extent by large variance penalty (figure 2, last column). 

\begin{figure}[h!]
	\centering
	\includegraphics[width=\linewidth]{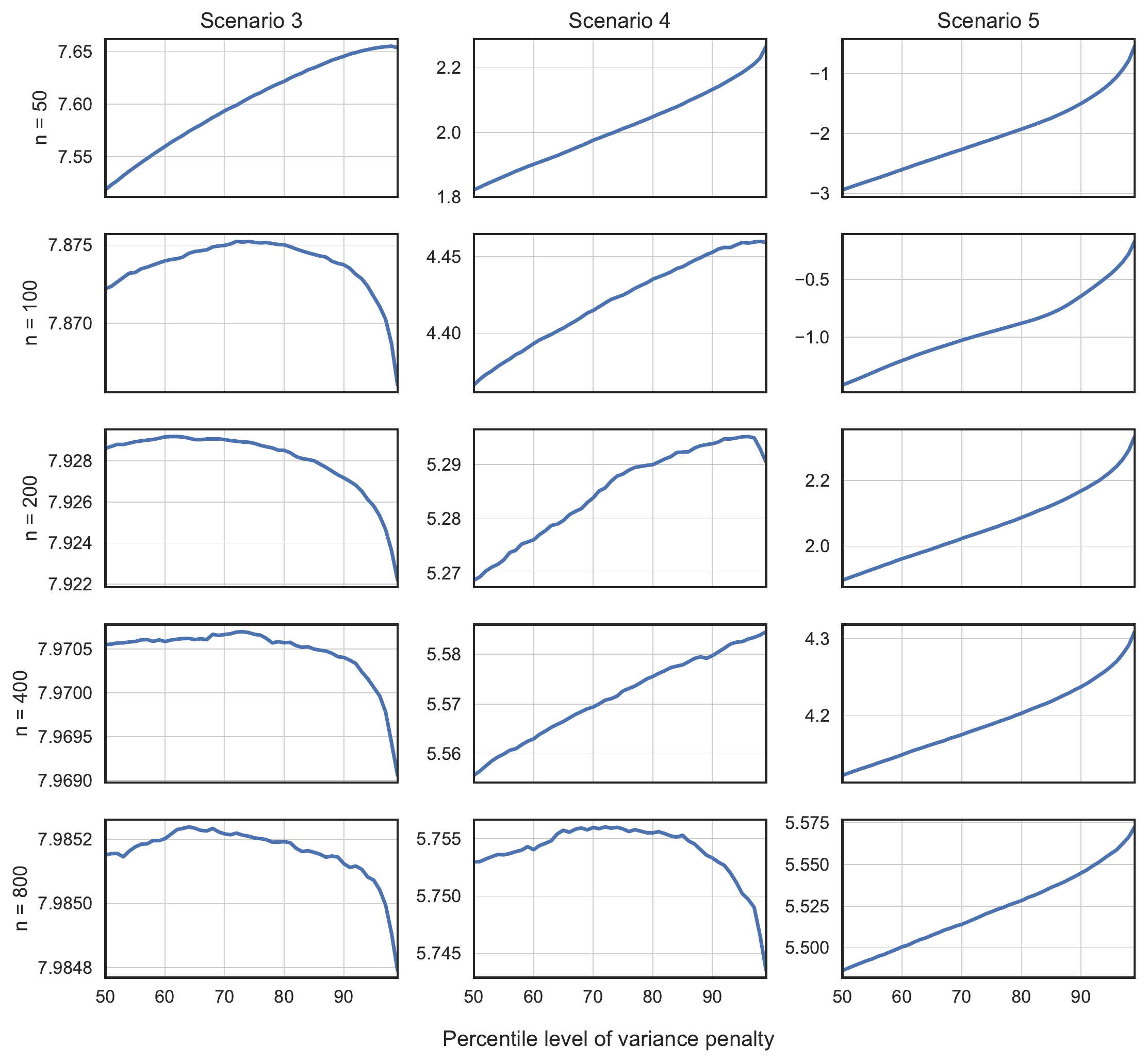}
	\caption{Dependence of average quality $ \hat{\mathcal{V}}(f) $ on $ s $-factor values (see (\ref{eq:objective})), which are represented as normal percentile levels, i.e. $ 95 $ percentile level means $ s = \Phi^{-1}(0.95) $, where $ \Phi $ is normal CDF.}
	\label{fig:sfactordynamics}
\end{figure}

LCSL performance dependence on the variance penalty coefficient, denoted by $ s $ in objective (\ref{eq:objective}), needs additional elaboration. We have varied $ s $ on standard normal quantile grid (from 0.5 to 0.99 quantile with a 0.01 step) and observed a sustainable relationship between penalty value and method performance (figure \ref{fig:sfactordynamics}). 

The first interesting effect, equally present in all scenarios, is that strong variance penalty always improves upon mean prediction given very small train data (positive trend on every plot in the first row of figure \ref{fig:sfactordynamics}). 

The second observation is that increasing train sample size in randomized studies reduces the optimal variance penalty (first two columns in figure \ref{fig:sfactordynamics}). This observation could be explained by increase of model accuracy and, as a consequence, diminished need in variance penalty. Note that for more complex setup (scenario 2) optimal penalty shrinks slower than for a simpler one (scenario 1). However, we also report a consistent small improvement in average $ \hat{ \mathcal{V}}(f) $ irrespective of train data size for small values of penalty (figure \ref{fig:sfactordynamics}). This effect can be viewed as a regularization, preventing the model of too assertive extrapolation. 

The last noteworthy remark concerns observational study (last column of figure \ref{fig:sfactordynamics}). In such setting we observe that large variance penalty increases performance irrespective of train sample size. 
That is because data generation policy ($ A \sim \mathcal{N}( f^{opt}_4(\C) , 0.5) $) is by design close to optimal. Thus, in observational studies the value of variance penalty should be treated as a proxy of prior knowledge about the degree of data generation policy optimality.

We conclude that different setups need to treat variance penalty differently. Randomized studies benefit most when treating variance penalty as a regularization. In contrast, observational studies should treat variance penalty as a means for explicit reasoning about degree of data generation policy optimality.

\clearpage

\section{Conclusions} 
We have presented an \emph{indirect} approach to IDR estimation which is based on Gaussian process regression. 

In spite of obvious difficulties of \emph{indirect} IDR estimation, we have shown that it is possible to reduce such an intractable problem to a univariate global optimization setting which is ubiquitous in Bayesian optimization and active learning.

We have discussed the advantages and disadvantages of the proposed method including explicit probability reasoning about potential outcome and computational challenges arising as a consequence of using flexible nonparametric model. 

We have also reported various simulation results which not only show that our method is highly competitive with other algorithms of ITR estimation, but uncover unexpected variance penalty interpretations. 

\section{Future work}

As was previously noted, ITR estimation shares many similarities with policy search problem arising in reinforcement learning (RL). This suggests researches dealing with ITR and DTR (dynamic treatment regime) estimation could get inspiration from already existing methods in reinforcement learning world.

For example, it would be very interesting to see an adaptation of reward weighted regression \citep{peters2007reinforcement} to \emph{direct} DTR estimation. Also, an extension of the proposed LCSL method to multi-stage treatment assignments would be of a great interest. Such extension could probably inherit some of the useful ideas from data-efficient RL algorithm called PILCO \citep{deisenroth2011pilco}.

Diffusion of statistical science and reinforcement learning may bring benefits to all parties and thus is highly desirable.

\bibliography{jsm_17}
\end{document}